\documentclass[12pt]{article}

\usepackage{epsfig}
\usepackage{amssymb}
\newcommand{\be}{\begin{equation}}
\newcommand{\ee}{\end{equation}}
\newcommand{\ba}{\begin{eqnarray}}
\newcommand{\ea}{\end{eqnarray}}
\begin{document}
\title{Determination of Resonances by the Optimized Spectral Approach}
\author{Arkadiusz Kuro\'s, Przemys\l aw Ko\'scik, Anna Okopi\'nska\\
Institute of Physics, Jan Kochanowski University\\
 \'Swi\c{e}tokrzyska 15, 25-406 Kielce, Poland}
 \maketitle
\begin{abstract}
The Rayleigh-Ritz procedure for determining bound-states of the Schr\"{o}dinger equation relies on spectral representation of the solution as a linear combination of the basis functions. Several possible extensions of the method to resonance states have been considered in the literature. Here we propose the application of the optimized Rayleigh-Ritz method to this end. The method uses a basis of the functions containing adjustable nonlinear parameters, the values of which are fixed so as to make the trace of the variational matrix stationary. Generalization to resonances proceeds by allowing the parameters to be complex numbers. Using various basis sets, we demonstrate that the optimized Rayleigh-Ritz scheme with complex parameters provides an effective algorithm for the determination of both the energy and lifetime of the resonant states for various one-dimensional and spherically symmetric potentials. The method is computationally inexpensive since it does not require iterations or predetermined initial values. The convergence rate compares favorably to other approaches.
\end{abstract}

\section{Introduction}\label{sec:intro}
Resonance phenomena appear in many fields of quantum physics: from unstable elementary particles, to resonances in atomic or molecular systems and to collective excitations in the condensed phase. They are described as long-lived states of a system that have enough energy to undergo a decay process. Wave functions of resonant configurations resemble bound-states over a period of time, called the decaying lifetime, when they are captured in a small area of space. Likewise the bound states, they may be treated as the eigenstates of the Hamiltonian that are associated with the natural frequencies of the system. The difference lies in the complex character of the resonance eigenvalues which is related to the purely outgoing boundary condition they fulfil. Such an approach to resonances has been started by Gamow in the paper on $\alpha$-decay of radioactive nuclei \cite{Gamow} and further developed by Siegert \cite{Siegert}. It has been shown that the problem of non-square integrability of their eigenfunctions can be rigorously overcome by enlarging the functional space to a rigged Hilbert space \cite{riggedGelfand, riggedMaurin}. On the other hand, the complex scaling idea~\cite{BC} has enabled development of practical approaches where the resonant states are treated on the same footing as the bound-states~\cite{Reinhardt,Moiseyev}. One of the most practical methods is the Rayleigh-Ritz (RR) determination of the eigenstates of a complex-scaled Hamiltonian. 

In this work, we discuss the determination of the resonances by the optimized RR scheme~\cite{ao} which proved successful for bound-states~\cite{Amore,RRopty,coll}. The method uses a basis of functions with adjustable nonlinear parameters, the values  of which are fixed so as to make the trace of the RR matrix stationary. Generalization to resonances proceeds straightforwardly by allowing nonlinear parameters to be complex numbers. Through the study of several systems with different potentials we demonstrate the efficiency of our method in finding both the energy and lifetime of the resonant states.

The plan of our work is as follows. In section~\ref{sec:RR}, the optimized RR method is described and its extension to resonances is discussed. 
The calculations of the resonance parameters are presented in section~\ref{sec:one} for one-dimensional potentials, and in section~\ref{sec:radial} for the spherically symmetric case in $D$ dimensions. Section~\ref{sec:con} is devoted to conclusion.

\section{The method}\label{sec:RR}
The Schr\"{o}dinger equation
\begin{equation} \hat{H}\psi(x)= \varepsilon \psi(x),\label{eq:prob1}\end{equation}
in the vast majority of cases cannot be solved exactly and has to be dealt with approximately. One of the classical methods for calculating numerically accurate solutions is the linear Rayleigh-Ritz procedure. 
\subsection{Rayleigh-Ritz determination of bound-states}
The RR method for solving the Schr\"{o}dinger equation (\ref{eq:prob1}) is based on the variational principle which says that the Rayleigh quotient \begin{equation} R[\Phi]=\frac{<\Phi(x)|\hat{H}|\Phi(x)>}{<\Phi(x)|\Phi(x)>},\label{eq:roqq}\end{equation}
achieves minimum when $\Phi(x)$ fulfils Eq.(\ref{eq:prob1}) with boundary conditions $\psi(x \to \pm \infty)\to 0$. This allows determination of bound-state wave-functions by approximating them by finite linear combinations   
\begin{equation} \Phi(x)=\sum ^{M-1}_{j=0} c_j \phi^{A}_j(x),\label{eq:baza}\end{equation} where the functions $\phi^{A}_j(x)$are taken from an orthonormal basis in the function space. 
The variational principle yields the matrix equation for the linear parameters $c_j$ in the form
\begin{equation}
\sum ^{M-1}_{j=0} (H^{A}_{jm} - \varepsilon \delta_{jm})c_j = 0,~m=0,1,...,M-1,
\label{eq:new}
\end{equation}
where the matrix elements of the Hamiltonian are \begin{equation}H^{A}_{jm}= \langle \phi^A_j |\hat{H}| \phi^A_m \rangle=\int_{-\infty}^{\infty}\phi^A_j(x)\hat{H}\phi^A_m(x)dx. \label{eq:real}
\end{equation}
Diagonalisation of the RR matrix provides $M$th order approximations to the $M$ of the lowest energy states. Systematically increasing the matrix dimension $M$, we increase the number of determined bound-states, wherein their approximate energies approach the exact results from above. The convergence of the method depends heavily on the choice of the basis set $\{\phi^{A}_j(x), j=0,1,...\}$. It appears advantageous to make the functions of the basis adaptable to the problem under study by allowing their dependence on nonlinear parameters ($A$), the values of which can be conveniently adjusted in each order approximation. To ensure a fast convergence of the particular eigenvalue (usually the ground-state energy), the values of nonlinear parameters are chosen by the trial and error or determined in numerically demanding optimization procedures \cite{rych}. Another option, that does not need any starting values or iterations, is to fix the nonlinear parameters according to the principle of minimal sensitivity \cite{PMS}, i.e. so that the approximation to a physical quantity would depend as weakly as possible on infinitesimal changes of their values. In the optimized RR method, proposed by one of us~\cite{ao}, the sum of $M$ bound-state energies is chosen as the physical quantity, the $M$th order approximation to which is given by the trace of the RR matrix 
$TrH_{A}^{(M)}=\sum^{M-1}_{j=0}<\phi^{A}_j|\hat{H}|\phi^{A}_j>$. The stationarity of the trace requirement
\begin{equation} \frac{\delta}{\delta A} TrH_{A}^{(M)}=0,\label{eq:sta}\end{equation}
is used to fix the values of nonlinear parameters, and a diagonalization of the so optimized matrix determines a set of $M$ aproximate eigenstates which are mutually orthogonal. The method is computationally less demanding although its convergence for a particular state may be slower than that achieved with nonlinear parameters iteratively optimized for that state. In the case of bound-states, the effectiveness of our method has been demonstrated for various potentials and various basis sets \cite{ao,Amore,RRopty,coll}. Here we extend its application to resonant states.

\subsection{Rayleigh-Ritz determination of resonant states}
Experimentally, the resonances manifest themselves as sharp peaks in the collision cross sections which are well described by the two parameter Breit-Wigner formula. The resonance energy $E$ and the half-width of the peak $\Gamma$ may be related to the complex eigenvalues 
\begin{equation}\varepsilon=E-i\Gamma/2, \label{eq:enercms}\end{equation}
of the Schr\"{o}dinger equation
(\ref{eq:prob1}), allowing $\Gamma$ to be interpretated as the inverse of the resonance lifetime. In finite range potentials, the wave function of a resonant state exhibits an asymptotic behavior of the form
\begin{equation}\psi_{rez}(x\to \pm \infty ) \approx e^{\pm ik_{rez}x} , \label{eq:asy1}\end{equation}
where
\begin{equation}k_{rez}= |k_{rez}|e^{-i\alpha_{rez}}, \label{eq:krez}\end{equation}
with $0<\alpha_{rez}<\pi /2$, as corresponds to the position of wave vector $k_{rez}$ in the fourth quarter of the complex plane. Since the wave function $\psi_{rez}(x)$ diverges exponentially, the Hamiltonian is not hermitian and its complex eigenvalues are hidden on a higher Riemann sheet of the complex energy plane. 

\subsubsection{Complex Scaling}
The complex scaling transformation 
\begin{equation}U(\theta): x \mapsto x e^{i\theta}, \label{eq:tran}\end{equation}
allows the treatment of resonant states in analogy to bound-states.
The corresponding complex-rotated Hamiltonian 
\begin{equation} \hat{H}_{\theta}=U(\theta) \hat{H}U^{-1}(\theta),\label{eq:hamt}\end{equation}
satisfies the eigenequation 
\begin{equation} \hat{H}_{\theta} \psi_{\theta}(x)=\varepsilon_{\theta} \psi_{\theta}(x),\label{eq:rsr}\end{equation}
and its spectrum is described by the Basley-Combes theorem. For dilatation analytic potentials~\cite{BC}, among which are the Coulomb and Yukawa potentials in addition to the finite range ones, the theorem states that the real bound-state eigenvalues, the complex resonance eigenvalues and the thresholds are the same as those of the original Hamiltonian, but the eigenvalues of the continuous spectrum are rotated about the thresholds by an angle $2\theta$ into the lower energy half-plane, exposing complex resonance eigenvalues. The complex scaling transformation (\ref{eq:tran}) turns the function $\psi_{rez}(x)$ into a normalizable one, if the real parameter $\theta$ is such that $0<\theta-\alpha_{rez}<\pi /2$. In this case, the resonances can be determined as the eigenstates of the non-hermitian Hamiltonian $\hat{H}_{\theta}$ by using bound-state-like strategies. Here we use the optimized RR method with a complex basis to this end. 

\subsubsection{Complex Basis}
Since the resonance eigenvalues are complex numbers, their spectrum is determined by stationarity rather than minimization condition. This requires that the Rayleigh quotient
\begin{equation} R[\Phi]=\frac{<\Phi(x)|\hat{H}_{\theta}|\Phi(x)>}{<\Phi(x)|\Phi(x)>},\label{eq:roqq}\end{equation}
be stationary at the square integrable solutions of the complex rotated Hamiltonian~(\ref{eq:hamt}). With the solution $\Phi(x)$ approximated by a finite linear combination of the real functions (\ref{eq:baza}), the same secular equation is obtained as for bound states (\ref{eq:new}) but
with the matrix element $H^A_{jm}$ replaced by 
\begin{equation}
H^{A,\theta}_{jm}=\langle \phi^A_j |\hat{H}_{\theta}| \phi^A_m \rangle=\int_{-\infty}^{\infty}\phi^A_j(x)\hat{H}_{\theta}\phi^A_m(x)dx.
\end{equation}
It has been observed~\cite{ComplexBasis} that changing the variable $x$ to $x e^{-i\theta}$ and using Cauchy's theorem to distort the integration contour back to the real axis, the matrix elements turn into
\begin{equation}
H^{A,\theta}_{jm}=\int_{-\infty}^{\infty}\phi^A_j(xe^{-i\theta})\hat{H}\phi^A_m(xe^{-i\theta})dx.\label{scale}
\end{equation}
The complex scaling is thus equivalent to working with original Hamiltonian and using the basis functions with coordinates rescaled with $e^{-i\theta}$ factor~\cite{Reinhardt,ComplexBasis}. Note, however, that instead of the ordinary scalar product of the Hilbert space  $<f|g>= \int_{-\infty}^{\infty} f^{*}(x)g(x) dx$, the c-scalar product
$(f|g)= \int_{-\infty}^{\infty} f(x)g(x) dx$ is to be used in the complex basis approach~\cite{nm}. The advantage of the complex basis approach is that it applies also for non-dilatation analitic potentials. Moreover, generalization to many-body systems allows introducing different scaling parameter for each degree-of-freedom, which makes the method more flexible than that of the complex scaled Hamiltonian. In the case when the nonlinear parameter $A$ is the scale parameter, so that \be \phi^{A}_j(x)={1\over \sqrt{A}}\phi_{j}\left({x\over A}\right),\ee the RR matrix element (\ref{scale}) may be written as
\begin{equation}
H^{\alpha}_{jm}=\int_{-\infty}^{\infty}\phi^\alpha_j(x)\hat{H}\phi^\alpha_m(x)dx,\label{Hcomplex}
\end{equation}
where $\alpha=Ae^{i\theta}$ and may be simply obtained by replacing the real parameter $A$ in (\ref{eq:real}) by a complex parameter $\alpha$. 

\subsubsection{Optimized RR method for resonances} 
In this work we generalize the optimized RR method~\cite{ao} to the case of resonant states by using the complex basis approach. Dealing with the unscaled Hamiltonian $\hat{H}$, we choose a basis set of real functions $\{\phi^{\alpha}_j(x), j=0,1,...\}$ so that the matrix elements $<\phi^{\alpha}_m|\hat{H}|\phi^{\alpha}_n>$ are given by analitic expressions. In the $M$-th order calculation, we fix the value of the parameter $\alpha$ to be equal to the complex solution of the stationarity of the trace condition (\ref{eq:sta}) and after a single diagonalization of the $M$-dimensional symmetric complex matrix we obtain a set of $M$ approximate eigenvalues. The number of calculated eigenvalues and their accuracy may be increased by increasing $M$, which permits quantification of the precision of the obtained results. In order to demonstrate the effectiveness of our method, we consider several very different systems described by one-dimensional and radial Schr\"{o}dinger equations, using various basis sets of functions with adjustable scale parameters. In some cases it turns out to be advantageous to introduce additional parameters that are not the scaling parameters.  

\section{One-dimensional Schr\"{o}dinger equations}\label{sec:one}
\subsection{Harmonic oscillator basis: $RR_{opt}^{\Omega}$ method}
First, we consider the case of reflection symmetric potentials $V(x)=V(-x)$. We use the $RR_{opt}^{\Omega}$ method with the basis of the harmonic oscillator (HO) eigenfunctions 
\begin{equation}  \phi_j^{\Omega}(x)=\left(\frac{\sqrt{\Omega}}{\sqrt{\pi}2^j j!}\right)^{1/2} H_j (\sqrt{\Omega}x) e^{-\frac{\Omega x^2}{2}}, \label{eq:one}\end{equation}
where the square root of the oscillator frequency $\Omega$ plays a role of an inverse scaling parameter. Due to reflection symmetry, the even- (odd-) parity states may be obtained by diagonalization of the Hamiltonian matrix in the basis of the first $M$ even (odd) functions, which substantially reduces the computational cost. 
\subsubsection{Quartic resonance potential} 
\begin{figure} [h!]

\begin{center}
\scalebox{0.9}{\includegraphics{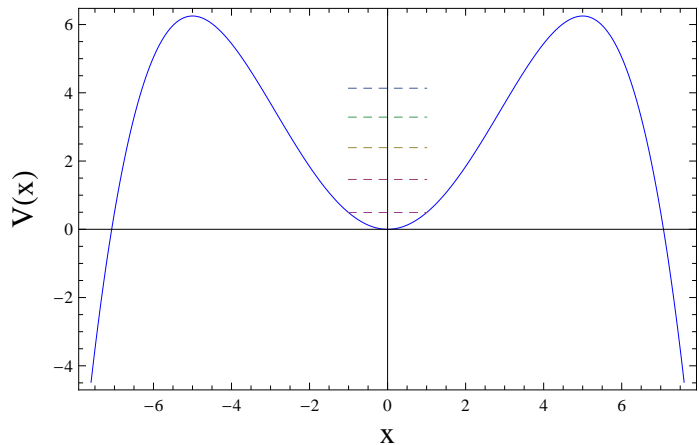}}
 \caption{The first five resonance's energies $Re\varepsilon$ for the potential $V(x)=\frac{1}{2}x^2-0.01 x^4$.}
 \label{fig:quartic}
 \end{center}
\end{figure}

A simple example of a resonant system is the inverse quartic anharmonic oscillator with a Hamiltonian
\begin{equation}
    \hat{H}=-\frac{1}{2}\frac{d^2}{dx^2}+ V(x)=-\frac{1}{2}\frac{d^2}{dx^2}+ \frac{1}{2}x^2 -\frac{\lambda}{2} x^4,
    \label{eq:hl1}
\end{equation}
where the units $\hbar=1$ and $m=1$ are used. In the case of $\lambda>0$, the potential $V(x)$ is not bounded from below and the system possess only resonant states, the lowest energies of which are marked on Fig.~\ref{fig:quartic} for $\lambda = 0.02$. The lifetime of the resonances increases with decreasing $\lambda$, and the $RR_{opt}^{\Omega}$ calculations require using larger basis sets to obtain a satisfactory accuracy. Our numerical approximations to $\epsilon_0$ and $\epsilon_2$ for $\lambda = 0.02$, which is the most demanding case considered in the literature \cite{ferSiegert}, are presented in table \ref{eq:tab1} as a function of the dimension of the RR matrix $M$. Here and in the following tables the results are accurate to the number of figures shown. The best accuracy is obtained for the lowest resonance, and for all states it quickly improves with increasing $M$. We can see that the literature results of the Riccati-Pad\'e method \cite{ferSiegert} are reproduced with RR matrix of dimension $M=25$; for larger $M$, more accurate values are easily obtained. The imaginary part of $\epsilon_2$ is larger than that of $\epsilon_0$, and we observed its further increase for higher resonances, which confirms that the lifetime decreases with increasing resonance energy. 

\begin{table} [h!]
\begin{center}
\scriptsize
\begin{tabular}{ccll}
\hline
$M$&$\Omega_{opt}$& $E_0$ & $\Gamma_0$\\
&&$E_2$&$\Gamma_2$\\
\hline
20&0.723 - 0.754 I&0.4922138348826277005 &$5.109*10^{-14}$\\
&&2.393167523963263&$2.842626*10^{-9}$\\
25&0.759 - 0.853 I&0.49221383488262770042136&$5.109394888*10^{-14}$\\
&&2.39316752396326281772&$2.84262607840*10^{-9} $\\
30&0.791 - 0.937 I&0.4922138348826277004213624190&$5.10939488839463*10^{-14}$\\
&&2.393167523963262817721830&$2.8426260783914863*10^{-9}$ \\
35&0.821 - 1.010 I&0.49221383488262770042136241897612033&$5.109394888394627276*10^{-14}$\\
&&2.39316752396326281772183026343&$2.842626078391486200557*10^{-9}$\\
\hline
\end{tabular}
\caption{Optimal values of the parameter $\Omega$ and the energies and widths of the two lowest even resonances for the Hamiltonian (\ref{eq:hl1}) with $\lambda =0.02$ calculated by the $RR_{opt}^{\Omega}$ method.}
\label{eq:tab1}
\end{center}
\end{table}

\subsubsection{Triple-well oscillator}
The sextic oscillator Hamiltonian
\begin{equation}
\hat{H}=-\frac{1}{2}\frac{d^2}{dx^2}+ \frac{1}{2}x^2 -g^2x^4+\frac{g^4}{2} x^6,
    \label{eq:hx6}
\end{equation}
where the triple-well potential $V(x)=\frac{1}{2}x^2 -g^2x^4+\frac{g^4}{2} x^6$ is bounded from below 
and increases to infinity at $|x|\to \infty$, describes an interesting system which supports only bound states. It has been shown~\cite{BGG,irregular} that the complex scaling transformation turns the asymptotically divergent solutions of this problem into square integrable ones that are associated with complex eigenvalues which describe the rates of tunneling between the potential wells. The eigenvalues $\epsilon_0$ and $\epsilon_4$ determined with the $RR_{opt}^{\Omega}$ method for various matrix dimensions $M$ are presented in table \ref{eq:tabg1} for $g=0.08$ and in table \ref{eq:tabg2} for $g=0.3$. We observe that with increasing $g$, the imaginary part of the eigenvalue grows, i.e. the resonance lifetime decreases, and the accuracy of the method improves. Comparison with the best published results for $\varepsilon_0$, obtained by the Ricatti-Pad\'e method \cite{ferSiegert}, shows that in the most unfavorable case of $g=0.08$ we attain the same level of accuracy with the matrix of dimension $M=60$, while $M=30$ is sufficient in the case of $g=0.3$.

\begin{table} [h!]
\begin{center}
\scriptsize
\begin{tabular}{ccll}
\hline
$M$ &$\Omega_{opt}$& $E_0$& $\Gamma_0$\\
&& $E_4$& $\Gamma_4$\\
\hline
40&0.5161 - 0.8673 I&0.4951282297707530015692954656&$10^{-28}$\\
&&4.28962031383623330069490&$4*10^{-20}$\\
50&0.5162 - 1.0001 I&0.4951282297707530015692954656874446&$1.2*10^{-32}$\\
&&4.289620313836233300694904739&$4.180046*10^{-20}$\\
60&0.5163 - 1.1171 I&0.4951282297707530015692954656874446179669&$1.16994174*10^{-32}$\\
&&4.2896203138362333006949047388180616&$4.1800456118133*10^{-20}$\\
70&0.5163 - 1.2230 I&0.495128229770753001569295465687444617966881040&$1.1699417439855*10^{-32}$\\
&&4.2896203138362333006949047388180616237064&$4.1800456118132521255*10^{-20}$\\
\hline
\end{tabular}
\caption{Optimal values of the parameter $\Omega$ and the energies and widths of the even resonances for the Hamiltonian (\ref{eq:hx6}) with $g =0.08$ calculated by the $RR_{opt}^{\Omega}$ method.}
\label{eq:tabg1}
\end{center}
\end{table}

\begin{table} [h!]
\begin{center}
\scriptsize
\begin{tabular}{ccll}
\hline
$M$ &$\Omega_{opt}$& $E_0$& $\Gamma_0$\\
&& $E_4$& $\Gamma_4$\\
\hline
10&0.5159 - 1.7799 I&0.40780&0.0294002\\
&&2.6095&4.7968\\
20&0.5163 - 2.5792 I&0.4078039790737&0.02940021689214\\
&&2.6094307234&4.79672853029\\
30&0.5163 - 3.1840 I&0.40780397907366957146&0.029400216892153485663\\
&&2.60943072337167570&4.79672853029023136\\
40&0.5164 - 3.6909 I&0.40780397907366957146548001&0.029400216892153485662928541\\
&&2.60943072337167569736879988&4.796728530290231359778820\\
50&0.5164 - 4.1363 I&0.40780397907366957146548000596165423&0.02940021689215348566292854082324361\\
&&2.60943072337167569736879987097172&4.7967285302902313597788200335659\\
\hline
\end{tabular}
\caption{Same as in Table \ref{eq:tabg1} but for $g =0.3$.}
\label{eq:tabg2}
\end{center}
\end{table}

\subsection{Shifted harmonic oscillator basis: $RR_{opt}^{\Omega,t}$ method}
Discussing problems with the resonant potential that is not symmeric about the origin, it is advantegeous to use $RR_{opt}^{\Omega, t}$ method with additional complex parameter $t$ that shifts the argument of the HO basis functions 
\begin{equation}
     \phi_j^{\Omega, t}(x)=\left(\frac{\sqrt{\Omega}}{\sqrt{\pi}2^j j!}\right)^{1/2} H_j \left(\sqrt{\Omega}(x-t)\right) e^{-\frac{\Omega (x-t)^2}{2}}.
     \label{eq:bazat}
\end{equation}
We apply the above basis to determine the spectrum of the cubic anharmonic Hamiltonian 
\begin{equation}
    \hat{H}=-\frac{1}{2} \frac{d^2}{dx^2}+\frac{1}{2} x^2 + \gamma x^3,
    \label{eq:hxc2}
\end{equation}
for an exemplary value of $\gamma=0.1$. In table \ref{eq:tabxc1}, the complex energies of the lowest resonant state are presented  with the corresponding optimal values of the nonlinear parameters $\Omega_{opt}$ and $t_{opt}$ for various dimensions $M$ of the RR matrix. The best previously published results \cite{k3} are reproduced with the RR matrix of dimension $M=40$. By increasing $M$, we easily obtain more accurate eigenvalues. It is interesting to note that although the shift parameter is not a scale parameter, its value determined by the stationarity of the trace condition turns out to be complex and this appears crucial for a fast convergence of the optimized RR scheme.
\begin{table} [h!]
\begin{center}
\scriptsize
\begin{tabular}{cccll}
\hline
M & $t_{opt}$ &$\Omega_{opt}$ &$E_0$& $\Gamma_0$ \\
\hline
20 &-0.67 - 2.26 I&1.02 - 0.66 I&0.48432&0.0000161\\
30 &-0.54 - 2.78 I&1.11 - 0.75 I&0.48431599700&0.0000161204\\
40 &-0.43 - 3.20 I&1.18 - 0.81 I&0.484315997004117&0.000016120419000\\
50 &-0.34 - 3.55 I&1.24 - 0.86 I&0.4843159970041175430&0.00001612041900013357\\
60 &-0.25 - 3.86 I&1.29 - 0.90 I&0.484315997004117543023&0.0000161204190001335639\\
\hline
\end{tabular}
\caption{Optimal values of nonlinear parameters and the energy and width of the lowest resonance state for the Hamiltonian (\ref{eq:hxc2}) with $\gamma=0.1$ calculated by the $RR_{opt}^{\Omega,t}$ method.}
\label{eq:tabxc1}
\end{center}
\end{table}

\subsection{Trigonometric basis: $RR_{opt}^{L}$ method}
Another convenient basis is provided by the set of trigonometric (TRIG) functions that satisfy the Dirichlet boundary condition at $x=\pm L$, where the width of the box $L$ serves as a scaling parameter. The even functions are of the form
\begin{equation} \phi_{j}^L(x)={1\over\sqrt{L}}\cos\left[\left(j+\frac{1}{2}\right)\frac{\pi x}{L}\right], \label{eq:te} \end{equation}
and the odd ones are given by
\begin{equation} \phi_{j}^L(x)={1\over\sqrt{L}}\sin\left[(j+1)\frac{\pi x}{L}\right]. \label{eq:to} \end{equation}
We use the TRIG basis to find the eigenvalues of the Hamiltonian with an inverted Gaussian potential with quartic perturbation
\begin{equation}
    \hat{H}=-\frac{1}{2}\frac{d^2}{dx^2}  -5e^{-0.1 x^2}-\frac{\lambda}{2} x^4.
    \label{eq:hex3}
\end{equation}
\begin{table} [h!]
\begin{center}
\scriptsize
\begin{tabular}{ccll}
\hline
$M$&$L_{opt}$& $E_0$ & $\Gamma_0$\\
&&$E_1$&$\Gamma_1$\\
\hline
20&5.114 + 2.888 I&-4.5665655093777188&0.0177068941054286\\
&5.155 + 2.915 I&-3.8381035089108826&0.2743215904678811\\
30&5.840 + 3.348 I&-4.5665655093777188168702314&0.0177068941054286198302607\\
&5.872 + 3.367 I&-3.8381035089108826586063027&0.27432159046788112432336\\
40&6.427 + 3.699 I&-4.5665655093777188168702313367733584&0.01770689410542861983026074495278\\
&6.454 + 3.715 I&-3.838103508910882658606302626356831&0.274321590467881124323361759999012\\
50&6.924 + 3.989 I&-4.5665655093777188168702313367733583615406723&0.017706894105428619830260744952784024460349\\
&6.947 + 4.003 I&-3.83810350891088265860630262635683043046341&0.27432159046788112432336175999901239136805\\                  
\hline
\end{tabular}
\caption{Optimal values of the parameter $L$ and the energies and widths of the two lowest resonances for the Hamiltonian (\ref{eq:hex3}) with $\lambda =0.08$ calculated by the $RR_{opt}^{L}$ method.}
\label{eq:tabl1}
\end{center}
\end{table}

\begin{table} [h!]
\begin{center}
\scriptsize
\begin{tabular}{ccll}
\hline
$M$&$L_{opt}$& $E_0$ & $\Gamma_0$\\
&&$E_1$&$\Gamma_1$\\
\hline
20&7.202 + 4.098 I&-4.52348287219&$2.20682*10^{-7} $\\
&7.264 + 4.135 I&-3.62331693435&0.00006769098\\
30&8.261 + 4.726 I&-4.5234828721860057186&$2.20682214291*10^{-7} $\\
&8.306 + 4.753 I&-3.6233169343531338&0.0000676909876076569\\
40&9.089 + 5.218 I&-4.52348287218600571860826015&$2.20682214290912985423*10^{-7} $\\
&9.126 + 5.240 I&-3.62331693435313378989538384&0.0000676909876076569149953\\
50&9.788 + 5.630 I&-4.5234828721860057186082601443634245&$2.2068221429091298542221588*10^{-7}$\\
&9.821 + 5.649 I&-3.62331693435313378989538384275228&0.00006769098760765691499530503897\\                
\hline
\end{tabular}
\caption{Same as Table \ref{eq:tabl1}, but for $\lambda =0.01$}
\label{eq:tabl1a}
\end{center}
\end{table}

\noindent The energies and widths presented in Tables \ref{eq:tabl1} and \ref{eq:tabl1a} show that already with matrices of dimension $M=20$ we reach the accuarcy of the literature results \cite{gauss}. We observe that the energy of the resonant state does not change much  with decreasing $\lambda$, while its width decreases rapidly to zero, switching to the bound state at $\lambda=0$.

\section{Radial Schr\"{o}dinger equation in $D$-dimensional space}\label{sec:radial}
For spherically symmetric problems in $D$-dimensional space, the solution of the Schr\"{o}dinger equation factorizes into the angular part given by hyperspherical harmonics and the radial part $R(r)$ that fulfils 
\begin{equation}
   \left[-\frac{1}{2}\frac{d^2}{dr^2}-\frac{(D-1)}{2r}\frac{d}{dr}+\frac{l(l+D-2)}{2r^2}+V(r)\right]R(r)=ER(r).
    \label{eq:hradial}
\end{equation}
By substituting $R(r)=r^{(1-D)/2}u(r)$, the radial equation is brought to the one-dimensional Schr\"{o}dinger form
\begin{equation}
    \hat{H}_r u(r)=\left[-\frac{1}{2}\frac{d^2}{dr^2}+\frac{\Lambda(\Lambda+1)}{2r^2}+V(r)\right]u(r)=Eu(r),
    \label{eq:hr}
\end{equation}
where $\Lambda=l+D/2-3/2$.
\subsection{Radial harmonic oscillator basis: $RR_{opt}^{\Omega}$ method}
For determining the resonant spectrum of radial anharmonic oscillators, it is convenient to use the basis of the spherically symmetric HO eigenfunctions that are given by
\begin{equation} 
\phi_{j\Lambda}^{\Omega}(r)=\sqrt{\frac{2 j! \Omega^{\frac{3}{2}+\Lambda}}{\Gamma(j+\Lambda+\frac{3}{2})}}r^{\Lambda+1} e^{-\frac{r^2\Omega}{2}} L_j^{\Lambda+\frac{1}{2}}(r^2\Omega).\label{eq:baza2}
\end{equation} 
As an example, we consider the case of two dimensions, where $\Lambda=l-1/2$. We determine the resonances in the inverted Mexican hat potential with the radial Hamiltonian given by
\begin{equation}
    \hat{H}=-\frac{1}{2}\frac{d^2}{dr^2} +\frac{l^2-\frac{1}{4}}{2r^2}+ \frac{1}{2}r^2 - \frac{g}{2} r^4,
    \label{eq:hr}
\end{equation}
where $g>0$. The above Hamiltonian has been studied before in Ref.\cite{k4} by the RR method with the same basis (\ref{eq:baza2}), but with iteratively adjusting the value of $\Omega$ in each order so as to obtain the best convergence for a single selected state. Our results obtained with $\Omega$ being fixed a priori from the trace condition~(\ref{eq:sta}) are presented in Table \ref{eq:tabr1}. The comparison shows that our method automatically provides a fast convergence in determining a number of resonances with different quantum number $n$ in one run. 
\begin{table} [!ht]
\begin{center}
\scriptsize
\begin{tabular}{ccll}
\hline 
$M$&$\Omega_{opt}$&$E_0^0$&$\Gamma_0^0$\\
&&$E_0^1$&$\Gamma_0^1$\\
\hline 
10&0.8982 - 1.1917 I&0.856745041& 0.04543667 \\  
&0.9095 - 1.2172 I&1.56818293&0.34682442\\
15&1.0000 - 1.4142 I&0.85674504114583&0.0454366707026\\
&1.0090 - 1.4333 I&1.568182929694&0.34682442355763\\
20&1.0832 - 1.5874 I&0.8567450411458273172&0.0454366707025907851 \\
&1.0907 - 1.6029 I&1.5681829296936992519&0.346824423557637908916\\
25&1.1545 - 1.7316 I&0.8567450411458273172142177&0.0454366707025907850723740 \\
&1.1611 - 1.7448 I&1.568182929693699251881382&0.3468244235576379089157114\\
30&1.2174 - 1.8564 I&0.8567450411458273172142177179279& 0.04543667070259078507237390548 \\
&1.2233 - 1.8681 I&1.568182929693699251881381495503&0.346824423557637908915711315930\\
\hline 
\end{tabular}
\caption{Optimal values of the parameter $\Omega$ and the energies and widths of the lowest resonances with $l=0$ and $l=1$ for the Hamiltonian (\ref{eq:hr}) with $g=0.1$, calculated by the $RR_{opt}^{\Omega}$ method for increasing dimension $M$.}
\label{eq:tabr1}
\end{center}
\end{table}

\subsection{Radial trigonometric basis: $RR_{opt}^{L}$ method}
For studying problems with $\Lambda=0$, it appears convenient to use the basis of antisymmetric trigonometric functions 
\begin{equation} \phi_{j}^L(x)=\sqrt\frac{2}{L}\sin\left[(j+1)\frac{\pi x}{L}\right],\label{eq:to1} \end{equation}
where $L$ represents the confinement radius. We apply the above basis to calculate the resonant states of the Bardsley Hamiltonian~\cite{Bar} 
\begin{equation}
    \hat{H}=-\frac{1}{2}\frac{d^2}{dr^2} + V_{0}r^2 e^{-r}.
    \label{eq:ht1}
\end{equation}
\begin{figure} [h!]
\begin{center}
\scalebox{0.8}{\includegraphics{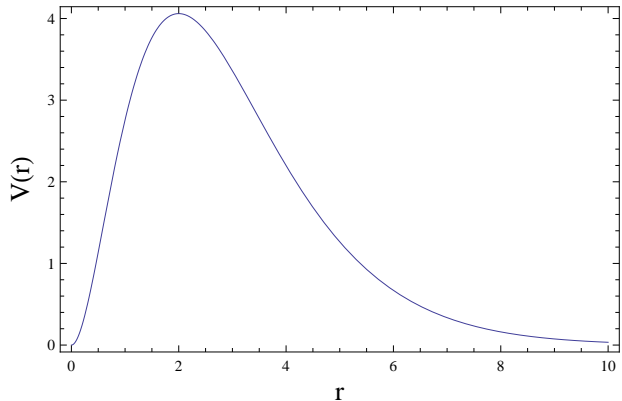}}
 \caption{Bardsley potential $V(r)=7.5r^2 e^{-r}$.} 
 \label{eq:p_nr12}
 \end{center}
\end{figure}
In Ref.\cite{rak}, the very accurate results obtained using the Jost-function method are presentes for nine lowest resonances. We show our results for the first and ninth state in Table \ref{eq:tabt1}. The convergence rate decreases with increasing resonance energy but with matrices of dimension $N=160$ we automatically reproduce all the results of Ref.\cite{rak}. Our results for the next resonance ($E_9$, $\Gamma_9$ are also presented in the Table. 
\begin{table} [!ht]
\begin{center}
\scriptsize
\begin{tabular}{ccll}
\hline 
$M$&$L_{opt}$&$E_0$&$\Gamma_0$\\
&&$E_8$&$\Gamma_8$\\
&&$E_9$&$\Gamma_9$\\
\hline 
100&-0.841+6.661I&3.4264&0.025549\\
&&-3.7  &40 \\
&& -6.9& 45\\
120&-1.099+6.804I&3.4263903&0.02554896\\
&& -3.75& 40.01\\
&& -6.801& 45.527\\
140&-1.319+6.920I&3.42639031015&0.02554896118\\
&& -3.754144 & 40.0090149\\
&& -6.80030& 45.5263\\
160&-1.511+7.018I&3.4263903101482&0.02554896118580\\
&& -3.754144122 &40.00901499 \\
&&-6.80030389   & 45.52631015\\
180&-1.680+7.102I&3.4263903101482505&0.025548961185791\\   
&&-3.754144122607 &40.0090149933 \\
&&-6.800303886379 &45.52631015000\\
\hline 
\end{tabular}
\caption{Optimal values of the parameter $L$ and the energies and widths of a few resonances for the Hamiltonian (\ref{eq:hr}) with $V_{0}=7.5$, calculated by the $RR_{opt}^{\Omega}$ method for increasing dimension $M$.}
\label{eq:tabt1}
\end{center}
\end{table}\newpage

\section{Conclusion}\label{sec:con}
We applied the optimized RR method with complex nonlinear parameters to determine the resonance states in various one-dimensional potentials. The expansion basis were adapted to the considered problems so that the RR matrix elements were given by analytic expressions. The values of nonlinear parameters were fixed by requiring that the trace of the truncated matrix be stationary. We have shown that the basis of the HO eigenfunctions with frequency $\Omega$ optimized by the stationarity of the trace condition is efficient in determining the resonance spectrum in anharmonic potentials. In the case of nonsymmetric about the origin potentials, an additional complex shift parameter $t$ has proved to be useful to obtain quick convergence. The trigonometric basis with optimized boundary period parameter $L$ appears convenient, especially in the case of potentials described by exponential functions. Effectiveness of a similar approach in determining resonances of the radial Schr\"odinger equation has been also shown. The advantage of the optimized RR scheme is that a set of resonances is determined automatically in one run without the necessity of specifying any starting value. The computational cost of our method is much lower than in the case of iterative optimization of nonlinear parameters. The method appears especially effective in the cases where only resonant modes exist. For the class of resonant potentials considered in the present work, it is highly competitive with existing methods, the results of which are easily recovered in our approach. 

\section{Acknowledgements}\label{ack} This work was partially supported by the ESF Human Capital Operational Programme grant 6/1/8.2.1./POKL/ 2009. One of the authors (AO) is deeply indebted to Jacek Karwowski for inspiring questions.\\

\end{document}